\documentclass[usenatbib, 12p]{article}
\usepackage{cancel, colortbl, booktabs}
\usepackage{comment, algorithm,algpseudocode, amsmath,epsfig,epsf, amssymb, amsfonts, latexsym, rotating,color}

\usepackage{amsmath, amsfonts, amssymb, natbib, graphicx, epsf, enumerate, color, multirow, latexsym, times, bm}

\def\bSig\mathbf{\Sigma}

\newcommand{\ba}{\begin{eqnarray}}
\newcommand{\ea}{\end{eqnarray}}
\newcommand{\bas}{\begin{eqnarray*}}
	\newcommand{\eas}{\end{eqnarray*}}
\newcommand{\ben}{\begin{enumerate}}
	\newcommand{\een}{\end{enumerate}}
\newcommand{\e}{E} % supposed to be italicized according to biometrika
\newcommand{\var}{\mbox{var}} % not italicized according to biometrika
\newcommand{\bit}{\begin{itemize}}
	\newcommand{\eit}{\end{itemize}}

\definecolor{lgray}{RGB}{150,150,150}	
\definecolor{dgreen}{RGB}{000,175,000} 
\definecolor{dred}{RGB}{125,000,000}

\newtheorem{lemma}{\sc Lemma}
\newtheorem{theorem}{\sc Theorem}

\newcommand{\convergeto}{ {\; \rightarrow \; }}

\def\T{{ \mathrm{\scriptscriptstyle T} }}

\title{Inference for case-control studies with incident and prevalent cases}
\author{\vspace{0.1in} Marlena Maziarz$^{a\dagger}$, Yukun Liu$^{b\dagger}$, Jing Qin$^c$ \& Ruth Pfeiffer $^{a\ddagger}$\\
\vspace{0.1in}
$^a$National Cancer Institute, National Institutes of Health, Rockville, MD, USA\\
\vspace{0.1in}
$^b$School of Statistics, East China Normal University, Shanghai, China \\
$^c$National Institute of Allergy and Infectious Diseases\\
\vspace{0.1in}
National Institutes of Health, Bethesda, MD, USA \\
\vspace{0.1in}
$^\dagger$ Equal contributions\\
$^\ddagger$ Corresponding author (E-mail: pfeiffer@mail.nih.gov)}

\begin{document}

\maketitle

\section*{Abstract}
We propose and study a fully efficient method to estimate associations of an exposure with disease incidence when both, incident cases and prevalent cases, i.e. individuals who were diagnosed with the disease at some prior time point and are alive at the time of sampling,  are included in a case-control study. 
We extend the exponential tilting model for the relationship between exposure and case status to accommodate two case groups, and  correct for the survival bias in the prevalent cases through a  tilting term that  depends on the parametric distribution of the backward time, i.e. the time from disease diagnosis to study enrollment. We construct an empirical likelihood that also incorporates   the observed backward times   for prevalent cases,  obtain efficient estimates of odds ratio parameters that relate exposure to disease incidence and propose a likelihood ratio test for model parameters that has a standard chi-squared distribution. We quantify the changes in efficiency of association parameters when incident cases are supplemented with, or replaced by, prevalent cases in simulations. We illustrate our methods by estimating associations of single nucleotide polymorphisms (SNPs) with breast cancer incidence in a  sample of controls and incident and prevalent  cases  from the U.S. Radiologic Technologists Health Study. \\

\emph{Keywords: Outcome dependent sampling, survival bias, empirical likelihood, exponential tilting model, density ratio model, length biased sampling.}

\section{Introduction \label{sec:intro}}

Case-control studies that compare the frequency of exposures in  incident cases to that in healthy individuals to assess  associations with risk of disease incidence are popular
 for  rare outcomes, as they are more economical than prospective cohorts. However, like all observational studies, case-control studies are also vulnerable to   biases   that result in distorted estimates of  exposures' associations with disease risk.  One of several possible biases 
, sometimes called   {\it  survival bias}, occurs 
when prevalent cases, i.e. individuals who were diagnosed with the disease at some prior time point and are alive at the time of sampling for the case-control study, are used in addition to, or instead of, individuals newly diagnosed with disease, namely incident cases.
If the exposure also impacts survival after disease onset, the estimated association of an exposure with disease incidence over- or underestimates the true association. 
This is a particularly serious problem for diseases, or outcomes, that are rapidly fatal, as survivors may comprise a very special subgroup of cases.
 
Many epidemiologic textbooks \citep[e.g.][p.~133]{schlesselman82} point out that simply including prevalent cases in case-control studies of rare diseases without 
any adjustment for the survival bias leads to   biased estimates of incidence odds ratios. 
While several authors have proposed approaches to correct for survival bias in the analysis of cohorts comprised of prevalent cases that are then followed to some failure event of interest (e.g. death) \citep[e.g.][]{cheng14a}, only one approach has been proposed to explicitly correct for survival bias when prevalent cases are compared with controls. 
\citet{begg87a} subtracted a bias term estimated from a survival model for the backward time from the log odds ratio estimates obtained from a standard logistic model fit to controls and prevalent cases.  A statistical approach to allow incorporating information from prevalent cases in addition to incident cases is thus needed to enhance inference based on case-control data for rare disease like cancer, where prevalent cases become more readily available due to improvements in treatment.

Our work was motived by a case-control study conducted within the U.S. Radiologic Technologists Study (USRTS) to assess the associations of single nucleotide polymorphisms (SNPs) with risk of female breast cancer \citep{bhatti08a}. The USRTS, initiated in 1982 
by the National Cancer Institute and other institutions to study radiation-related health effects from low-dose occupational radiation exposure, enrolled 146,022 radiologic technologists at baseline. Information on participants' characteristics, exposures and prior health outcomes was collected via several surveys conducted between 1984 and 2014, and blood sample collection for molecular studies began in 1999. 
 As the number of incident breast cancer cases with blood samples available for genetic analysis was limited, we developed methods that allow one to also include information on prevalent cases, i.e. women whose breast cancers were diagnosed prior to blood sample collection,  to obtain unbiased estimates of odds ratios for the associations of SNPs with breast cancer incidence.

Our work is based on the well known  result on the equivalence  between the logistic regression model for prospectively collected  data  and the exponential tilting, or density ratio model, for retrospectively collected data \citep{qin98a}. To accommodate data from incident cases, prevalent cases and controls, we discuss a three-sample exponential tilting density ratio model. For prevalent cases, in addition to covariate information, we observe their backward time, i.e. the time between disease diagnosis and sampling. We model the backward time distribution based on a parametric model for the survival time conditional on surviving to time of sampling (Section 2). 
In Section 3 we derive a semi-parametric likelihood that combines information from controls, incident and prevalent cases. We estimate log odds ratios for the associations between disease incidence and exposures, and parameters in the model for the backward time using empirical likelihood techniques, and derive the asymptotic properties of the estimates. In Section 4, we assess the performance of the method in simulations and study efficiency of the estimates when prevalent cases are used in addition to, or instead of, incident cases in a study under various scenarios. We illustrate the methods with data from the
motivating study on the association of breast cancer risk and SNPs among women sampled from the USRTS (Section 5), before closing with a discussion (Section 6).

\section{Semi-parametric model for case-control studies with incident and prevalent cases \label{sec:general-setup}}

\subsection{Background: exponential tilting model \label{sec:notation}}

Let $D$ denote the disease indicator, with $D = 1$ for individuals newly diagnosed with disease (incident cases) and 
$0$ for those without (controls), and $X$ is a vector of covariates. 
We assume that the association between $X$ and $D$ in the population is captured by the prospective logistic model
\begin{equation}
\label{pd1.x}
P(D=1\mathbin{|}X=x)=\frac{\exp(\alpha_0+x\beta)}{1+\exp(\alpha_0+x\beta)}
\end{equation}
where $\alpha_0$ denotes an intercept term, and $\beta$ the log odds ratio for the association of $X$ with $D$, the parameter of interest. 
In the general population, the marginal probability of disease is $\pi = P(D=1)=\int P(D=1\mathbin{|}x)f(x)dx$ where $f(x)=dF(x)/dx$ is the density of $X$, that is unspecified.

In a case-control study, independent samples of fixed sizes $n_0$ and $n_1$ are drawn from controls ($D=0$) and cases ($D=1$), respectively, and then information on the exposure $X$ is obtained. Due to the retrospective sampling, only the conditional densities $f_0(x) = f(x\mathbin{|}D=0)$ and $f_1(x) = f(x\mathbin{|} D=1)$ are observed. Using Bayes' rule, the prospective model in (\ref{pd1.x}), and letting $\alpha^* = \alpha_0 + \log\{(1-\pi)/\pi\}$, 
\begin{eqnarray}
\label{eq:tilt1}
f_1(x) &=& \frac{\exp(\alpha_0 + x\beta)}{1 + \exp(\alpha_0 + x\beta)}\frac{f(x)}{\pi} = f_0(x) \exp(\alpha_0 + x\beta) \left(\frac{1-\pi}{\pi}\right) =  f_0(x)\exp(\alpha^* + x\beta).
\end{eqnarray}
Model (\ref{eq:tilt1}) is called a two-sample exponential tilting model or density ratio model.

\citet{prentice79a} showed that ignoring the case-control sampling scheme and fitting the prospective model (\ref{pd1.x}) to the retrospectively ascertained exposure data yields consistent estimates of $\beta$ and the corresponding standard errors. 
\citet{qin98a} profiled out the baseline distribution  $f_0(x)$ in equation (\ref{eq:tilt1}) and derived a constrained empirical likelihood to estimate $\beta$ and the nuisance parameter $\alpha^*$.
We adapt this profile likelihood method  in the next section to  incorporate information on prevalent cases.

\subsection{Data and models for prevalent cases}
\begin{figure}
	\centerline{\includegraphics[scale = 0.6]{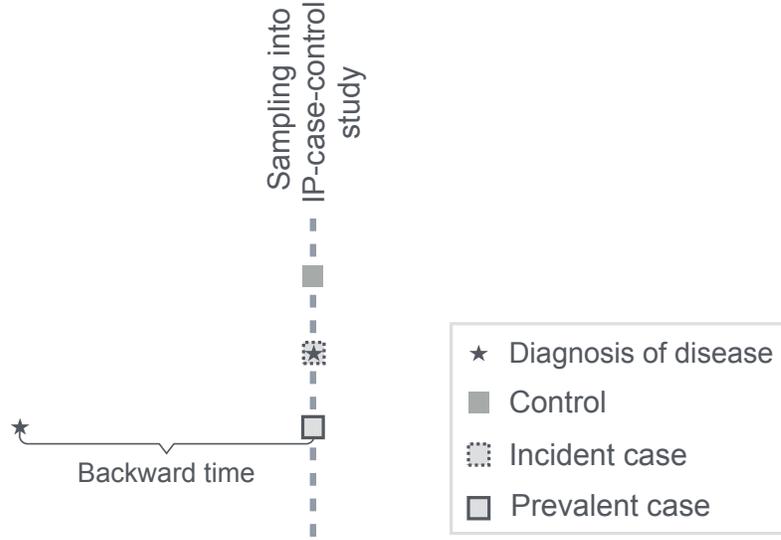}}
	\caption{Sampling scheme for the IP-case-control study design. For incident cases, disease  diagnosis (indicated by the star) occurs in the case-control sampling period. For prevalent cases, disease is diagnosed before sampling, and information on the backward time, i.e. the time between sampling and diagnosis, is also available. 
}
	\label{fig:prev.cc}
\end{figure}

We now assume that in addition to incident cases, on whom exposure information is ascertained at time of diagnosis, we also have information on 
exposures $X$ from prevalent cases, i.e. individuals who developed disease previously and are alive at the time of sample selection for the case-control study. To formalize the notion of a prevalent case, let $T$ denote the (unobserved) survival time from disease diagnosis to death, with a survival function $S(t) =P(T>t)$, and let $A$ denote the backward time, defined as the time between disease diagnosis and sampling. We only observe prevalent cases who are alive at the time of sampling, i.e. if $T > A$. The sampling scheme for incident cases, prevalent cases and controls is depicted in Figure \ref{fig:prev.cc}. If $X$ is related to survival, that is $S(t)=S(t\mathbin{|}x) =P(T>t\mathbin{|}x),$ simply combining prevalent with incident cases and fitting model (\ref{pd1.x}) to the data will lead to biased estimates of $\beta$.  In what follows we  assume that $S$ belongs to a known parametric family indexed by parameters $\gamma$ and use the notation
$S(.|x,\gamma)$. 

Using information on the covariates $X$ and the observed backward time, $A$, we now derive the likelihood contribution for the prevalent cases and extend the exponential tilting model in (\ref{eq:tilt1}) to obtain unbiased estimates of $\beta$. 
The joint distribution of $X$ and $A$ in the population of prevalent cases alive at the time of sampling is
\begin{eqnarray}
\label{pd2.x}
f(X=x, A = a \mathbin{|}D=1, T > A) = f(X=x\mathbin{|}D=1, T>A)f(A=a\mathbin{|}X = x,D=1, T>A). 
\end{eqnarray}

Assuming that the disease incidence is stationary over time, based on standard renewal process theory, we assume that $A$ has a uniform density on $[0, \xi]$, i.e. $f(a) = \xi^{-1} $ for $a \in [0, \xi]$ and zero otherwise,    for some  known value $\xi$.  Then using equation (\ref{eq:tilt1}) and Bayes' theorem, the density of the covariates $X$ for prevalent cases is
% \begin{eqnarray}
\begin{multline}
\label{eq:f2x}
f_2(X) = f(X=x|D=1, T>A) = \frac{P(T>A|X=x, D = 1) f(X=x | D = 1)}{P(T>A | D=1)} 
\\
= f_1(x) \frac{\int_0^{\xi}S(a|x, \gamma)da }{\int_{X}\int_0^{\xi}S(a|x, \gamma)da\, f_1(x) \,dx} 
= f_0(x) \exp\left\{\nu^* + x\beta + \log\mu(x, \gamma)\right\},
\end{multline}
where $\mu(x, \gamma)=\int_0^{\xi} S(a|x, \gamma)da$ and $\nu^* = \alpha^* -\log\{ \int_X \mu(x, \gamma) f_1(x) dx$\}.

The density of the covariates for the prevalent cases, $f_2(x)$ in (\ref{eq:f2x}) can thus also be expressed in terms of $f_0(x)$ and a parametric tilting term that, in addition to $\beta$ and an intercept, depends on the survival distribution $S$. Notice that when $S$ does not depend on $X$, the tilting term in (\ref{eq:f2x}) depends on $X$ only through $X\beta$, i.e. is the same as for the incident cases, but with a different intercept.
To derive the conditional density of $A$ in (\ref{pd2.x}), we use Bayes' theorem and the fact that $A$ is independent of both, $X$ and $T$, to obtain 
%\begin{multline}
\begin{eqnarray}
\label{fa.x2}
f_A(A = a| X=x, D = 1, T>A) &=&\frac{f(A = a) \,P(T > a \mathbin{|} X = x, D = 1)}{P(T > A \mathbin{|} X = x, D =1)} \nonumber \\
&=& \frac{\xi^{-1}S(a\mathbin{|}x, \gamma)}{\xi^{-1}\int_0^\xi S(a\mathbin{|}x, \gamma)} = \frac{S(a\mathbin{|}x, \gamma)}{\mu(x, \gamma)}, \mbox{ for } a \in [0,\xi]. 
%\end{multline}
\end{eqnarray}
For $ a \not \in [0,\xi]$, $f_A(A = a| X=x, D = 1, T>A)=0$.  
%From (\ref{fa.x2}) we see that while the support of $S$ can be infinite,  the conditional density of $A$ has support only on $[0,\xi]$. 
Any parametric survival model can be used to model $S$ in equation (\ref{fa.x2}) for the backward time $A$. We assume $S(t) = \exp(-\int_{0}^{t} h(s) ds)$ with hazard
$h(t) = h_0(t) \exp(x\zeta)$ where $h_0(t)$, the baseline hazard function, is modeled as a constant, Weibull or a piecewise-constant hazard.

\section{Semi-parametric likelihood and inference \label{sec:estimation}}

Let $(x_1,\ldots,x_{n_0})'$ denote the covariates for the $n_0$ controls,
$(x_{n_0+1},\ldots,x_{n_1})'$ the covariates for the $n_1$ incident cases and 
$(x_{n_0+n_1+1},\ldots,x_{N})'$ and $(a_{n_0+n_1+1}, \ldots, a_{N})^{\prime}$ 
the covariates and backward times for the $n_2$ prevalent cases, where $N=n_0+n_1+n_2$. Using the exponential tilting models in equations (\ref{eq:tilt1}) and (\ref{eq:f2x}), and the distribution for the backward time in (\ref{fa.x2}), the likelihood for the controls and the two case groups is
%{\footnotesize 
\begin{eqnarray}
\mathcal{L}
&=& \prod_{i=1}^{n_0}f_0(x_i) 
\prod_{i=n_0+1}^{n_0+n_1}f_0(x_i)\exp(\alpha^*+x_i \beta) \prod_{i=n_0+n_1+1}^{N}  f_0(x_i)\exp\{\nu^*+x_i\beta+\log\mu(x_i, \gamma)
\} \frac{S(a_i\mathbin{|}x_i, \gamma)}{\mu(x_i, \gamma)}  \nonumber \\ 
&=& \prod_{i=1}^N f_0(x_i) \prod_{i=n_0+1}^{n_0+n_1}\exp(\alpha^*+ x_i\beta)  \prod_{i=n_0+n_1+1}^{N}  \exp\{\nu^*+x_i\beta+\log\mu(x_i, \gamma)\} \frac{S(a_i\mathbin{|}x_i, \gamma)}{\mu(x_i, \gamma)}.
\label{fullL}
\end{eqnarray}

Similar to \cite{qin98a}, we estimate $p_i=f_0(x_i) = P(X=x_i), i=1,\ldots,N, $ empirically under the following constraints that ensure that $f_i, i=0,1,2$ are, in fact, distributions: \mbox{(1) $\sum_{i=1}^Np_i = 1, p_i \geq 0$}, (2) $\sum_{i=1}^N p_i\exp(\alpha^*+x_i\beta) = 1$, (3) $\sum_{i=1}^Np_i\exp\{\nu^*+x_i\beta+\log\mu(x_i, \gamma)\} = 1$. These constraints are accommodated via Lagrange multipliers in the log-likelihood. After maximizing the log-likelihood for $p_i$ subject to constraints (see Appendix 1), and letting $\alpha=\alpha^*+\log \left(n_1/n_0\right)$, $\nu =\nu^*+\log\left(n_2/n_0\right)$, the profile log-likelihood for the remaining parameters $\theta = (\alpha, \nu, \beta, \gamma)^\T$ is
\begin{multline}
\label{loglik}
\ell_p(\theta)
= -\sum_{i=1}^N\log\left[1+\exp(\alpha+x_i\beta)+\exp\{\nu+x_i\beta+\log\mu(x_i, \gamma)\}\right] \\
+ \sum_{i=n_0+1}^{n_0+n_1}(\alpha+x_i\beta)+\sum_{i=n_0+n_1+1}^N\left[\nu+x_i\beta+\log\mu(x_i, \gamma) + \log\left\{\frac{S(a_i\mathbin{|}x_i, \gamma)}{\mu(x_i, \gamma)}\right\}\right].
\end{multline}
We refer to the above likelihood as the {\it IP-case-control likelihood}.

Denote the maximum likelihood estimator of $\theta=(\alpha, \nu, \beta, \gamma)$ in (\ref{loglik}) by $\hat \theta = \arg\max_{\theta} \ell_p(\theta)$, and the true value by $\theta_0 = (\alpha_0, \nu_0, \beta_0, \gamma_0)^\T$. To derive the large-sample properties of $\hat \theta$, we first define a matrix $V$ that is important in the asymptotic behavior of $\hat \theta$. 

For ease of exposition, let $ w_1(x) = \exp(\alpha_0 + x \beta_0)$,
$w_2(x) = \exp\{\nu_0 + x \beta_0 +\log \mu(x,\gamma_0)\} $
and
$ \eta(x)=1+w_1(x) +w_2(x)$. 
Let $\e_0$ denote the expectation with respect to $dF_0(x)$, $\e_a$ the expectation with respect to $dF_A(a\mathbin{|}x)$, $Z^{\otimes 2} = ZZ^\T$ for any vector $Z$, 
and $\nabla_{\phi}$ denote the differentiation operator with respect to a generic parameter $\phi$.

We assume the following regularity conditions hold:
\begin{center}
	\begin{minipage}{15cm}  
		\ben
		\item[(C1)]
		$n_i/N \rightarrow \rho_i \in (0, 1)$ as $N\rightarrow \infty$ for $i=0, 1, 2$.
		
		\item[(C2)]
		For each $x$, both
		$\mu(x, \gamma)$ and $S(x, \gamma)$ have continuous first derivatives with respect to $\gamma$ in a neighborhood
		of $\gamma_0$.

\item[(C3)]
$
\e_0 \left[ \e_a\left\{ w_2(X)
\| \nabla_{\gamma } \log S(A\mathbin{|}X,\gamma) \|^2
\mid X \right\} \right] <\infty
$
and
$\e_0(\|X\|^2)$
is finite and positive definite.
\een
\end{minipage}
\end{center}

Condition (C1) requires that the sample sizes of the observed controls, incident cases and prevalent cases
grow at the same rate.
Note that $\int w_1(x) dF_0(x) = \rho_1\rho_0^{-1}$ and $\int w_2(x) dF_0(x) = \rho_2\rho_0^{-1}. $

Conditions (C2) and (C3) together guarantee that the matrix $ V \equiv (V_{ij})_{1\leq i, j\leq 4}$, where 
\begin{eqnarray*}
	V_{11} &=& \rho_0 \e_0\left\{ \frac{w_1^2(X) }{\eta(X)}\right\} -\rho_1, \; V_{12} = \rho_0 \e_0\left\{ \frac{w_1(X) w_2(X) }{\eta(X)}\right\} \; V_{13} = -\rho_0 \e_0\left\{ \frac{ w_1(X)X }{\eta(X)}\right\}, \\
	V_{14} &=& \rho_0 \e_0\left\{ \frac{ w_1(X)w_2(X) \nabla_{\gamma} \{ \log \mu(X,\gamma_0)\} }{\eta(X)}\right\}, \; V_{22} = \rho_0 \e_0\left\{ \frac{w_2^2(X) }{\eta(X)}\right\} -\rho_2,\\ 
	V_{23} &=& -\rho_0 \e_0\left\{ \frac{ w_2(X)X}{\eta(X)}\right\}, \; V_{24} = - \rho_0 \e_0\left\{ \frac{ \{1+w_1(X)\} w_2(X) \nabla_{\gamma} \{ \log \mu(X,\gamma_0)\} }{\eta(X)}\right\},\\
	V_{33} &= & - \rho_0 \e_0\left[ \frac{ \{ w_1(X)+w_2(X) \} XX^\T }{\eta(X)}\right], \; V_{34} = - \rho_0 \e_0\left[ \frac{ X w_2(X) \nabla_{\gamma} \{ \log \mu(X,\gamma_0)\}}{\eta(X)}\right], \\
	V_{44} &=& \rho_0 \e_0\left[ \frac{ w_2^2(X) }{\eta(X)} [\nabla_{\gamma} \{ \log \mu(X,\gamma_0)\}]^{\otimes 2}
	\right]- \rho_0 \e_0 \left\{ w_2(X) \e_a\left[ \{ \nabla_{\gamma } \log S(A\mathbin{|}X,\gamma_0) \}^{\otimes 2}
	\mathbin{|}X \right] \right\}.
\end{eqnarray*}
is well-defined. Condition (C3) ensures that all the moments in the definition of the $V_{ij}$'s are finite.

\begin{theorem}
	Assume regularity conditions (C1-C3) hold and that the matrix $V$ is positive definite.
	Then, as $N\rightarrow \infty$,
	\ben
	\item[(1)]
	$N^{1/2}(\hat \theta -\theta_0) \convergeto N(0, \Omega)$ in distribution,
	%where $\convergeto $ stands for ``convergence in distribution'', 
	$\Omega = V^{-1} \Sigma V^{-1}$, with $\Sigma$ defined in Supplementary Section 1; %\ref{proof-u};
	
	\item[(2)]
	the likelihood ratio
	$2\{ \sup_{\beta, \gamma}\ell_p(\beta, \gamma) - \ell_p(\beta_0, \gamma_0) \}
	\convergeto \chi_p^2$ in distribution, where $p$ is the dimension of $(\beta, \gamma)$;
	
	\item[(3)]
	the likelihood ratio for any sub-vector $\phi$ of $(\beta, \gamma) \convergeto 
	\chi_k^2$ in distribution, where $k$ is the dimension of $\phi$.
	
	\een
\end{theorem}
The proof of the Theorem is given in Appendix 2. 
Statement (3) implies that the standard $\chi^2$ asympotics  also hold for the likelihood ratio test restricted to the parameters $\beta$ of the model, which is of primary interest in many practical settings.

\section{Simulations}
We assessed the proposed model and estimating procedure in samples of realistic size, and characterized efficiency of estimates of the log odds ratio parameters $\beta$ when prevalent cases are used in addition to, or instead of, incident cases in a case-control study. 

\subsection{Data generation}

We generated data directly from the exponential tilting models 
(\ref{eq:tilt1}) and (\ref{eq:f2x}). 
For the controls, we simulated $n_0$ normally distributed covariates $(X_{01}, X_{02})' \sim N((0,0)', \Sigma)$, where for $ i = 1, 2$, $\Sigma_{ii} = 1,$ and $\Sigma_{ij} = \Sigma_{ji}= \rho$, $i \ne j$, with $\rho = -0.5, 0$ or $0.5$. For incident cases, we generated $n_1$ values $(X_{11}, X_{12})' \sim N(\Sigma(\beta_1, \beta_2)', \Sigma)$, where $(\beta_1, \beta_2)=(0, 0)$, $(1, 1),$ $(1, -1)$ or $(-1, -1)$.
To simulate exposure data $(X_1, X_2)$ for $n_2$ prevalent cases, we first generated a large set of values 
$(\tilde X_{k1}, \tilde X_{k2})' \sim N((0,0)', \Sigma)$, $k=1,\ldots, \tilde n_2$, where $\tilde n_2 \gg n_2$. For each $\tilde x_k = (\tilde x_{k1}, \tilde x_{k2})'$ we computed a weight $ \tilde w_2(\tilde x_k) = \exp\{\tilde x_k \beta+ \log \mu(\tilde x_k, \gamma)\}$, 
which is proportional to the tilt in equation (\ref{eq:f2x}), 
and then drew a sample of size $n_2$ with replacement, where each $\tilde x_k$ was sampled with probability $\tilde w_2(\tilde x_k)/\sum_j \tilde w_2(\tilde x_j)$.
As $P(\tilde X\mathbin{|}\tilde X \mbox{ sampled}) \propto \tilde w_2(\tilde X) f_0(\tilde X)$, the resulting sample has density $f_2$ as in equation (\ref{eq:f2x}).

For the survival distribution $S$, we assumed a proportional hazards model with a Weibull baseline hazard $h_0(t) = (\kappa_1/\kappa_2)(t/\kappa_2)^{(\kappa_1-1)}$ where $\kappa_1 >0$ and $\kappa_2>0$ are the shape and scale parameters, respectively, leading to $S(t\mathbin{|}x, \gamma) = \exp\left\{-(t/\kappa_2)^{\kappa_1} \exp(x\zeta) \right\}$, where $\zeta = (\zeta_1, \zeta_2)'$ are the parameters associated with covariate vectors $(X_1, X_2)_{N\times 2}$, with $(\zeta_1, \zeta_2) = (0, 0)$, $(1, 1),$ $(1, -1)$ or $(-1, -1)$. Then, $\mu(x, \gamma) = \Gamma(1/\kappa_1)/ (\kappa_1 \psi^{(1/\kappa_1)}) \left\{\Gamma^{-1}(1/\kappa_1)\int_0^{\psi\xi^{\kappa_1}}\exp(-u)\,u^{(1/\kappa_1-1)}\,du\right\}$ where $\psi = \kappa_2^{-\kappa_1}\exp(x\zeta)$, and the expression in the curly brackets is the cumulative distribution function of a Gamma distribution with shape parameter $\kappa_1^{-1}$ and scale parameter one, which can be evaluated using standard statistical software. 

In all simulations the  backward times for the prevalent cases were generated letting $\kappa_1=1$ to obtain a closed-form expression for $F_A$ in (\ref{fa.x2}). To obtain backward times, we first generated $U_i \sim \text{Unif}(0,1)$ and then computed $A_i = (1/\psi)[-\log\{1-U_i\,\psi\,\mu(x_i, \gamma)\}],$ $i=1,\ldots, n_2$. 

Estimates (Est) of the parameters, empirical standard deviations (SD$_{\text{emp}}$) of the estimates and standard deviations (SD$_{\text{asy}}$) were based on K=1000 replications for each parameter setting. SD$_{\text{asy}}=(\widehat{\Omega}/N)^{1/2}$ where $\widehat{\Omega}$ is the estimate of $\Omega$ in Theorem 1 scaled by the sample size.
We estimate this quantity as $\widehat{\Omega}/N = \sum_{i=1}^K(\widehat{V}^{-1}\widehat{\Sigma}\widehat{V}^{-1})/K$, where $\widehat{V}$ is the numerical estimate of the Hessian, and $\widehat{\Sigma}$ is obtained by computing the empirical variance of the scores for controls, incident and prevalent cases separately, and scaling each of the variance estimates by the respective sample size, to account for the retrospective sampling.

\subsection{Adding an increasing number of prevalent cases \label{sec:adding}}

We first examined the efficiency of the log odds ratio estimates $(\hat \beta_1, \hat \beta_2)$, and the estimates ($\hat \kappa_1, \hat \kappa_2, \hat \zeta_1,\hat \zeta_2$) of the parameters in the survival sub-model when $n_2 = 500$ or $1000$ prevalent cases were added to a study with $n_0 = 500$ controls and $n_1 = 500$ incident cases. 

Results for $\beta = (0, 0)$ and $\beta = (1, -1)$, both with $\zeta = (1, -1)$, $\rho = 0.5$ and $\xi = 25$ presented in Table \ref{table:1} show that the estimates of all parameters were virtually unbiased, and the asymptotic and empirical standard deviation estimates agreed well. As $n_2$ increased from 500 to 1000, the SD$_{\text{emp}}$ for $\widehat{\zeta}$'s decreased at the rate of $n_2^{-1/2}$, from 0.105 to 0.070 for $\beta= (0,0)$ and from 0.100 to 0.070 for $\beta = (1, -1)$. 

Efficiency results in terms of the ratio of the variance of $\hat \beta$ estimated using only the original $500$ incident cases and controls, compared to the variance of $\hat \beta$ when $n_2$ prevalent cases were added, for all combinations of $(\beta_1, \beta_2)$, $(\zeta_1, \zeta_2)$ and $\rho$ are shown in Figure 2. As the number $n_2$ of prevalent cases increased from 0 to 1000, efficiency gains under the null hypothesis, $\beta_1=\beta_2=0$, were modest and did not depend on the values of $(\zeta_1,\zeta_2)$ or $\rho$ (Figure 2a); the standard deviations (SD$_{\text{emp}}$) for $\widehat{\beta}$'s decreased slightly from 0.073 for $n_2=0$ to 0.069 for $n_2=1000$ (Table \ref{table:1}a). 

Efficiency gains were somewhat more noticeable for $\beta = (1, -1)$ and were greatest when $(\beta_1 ,\beta_2)$ had the same magnitude and signs as $(\zeta_1, \zeta_2)$, and $X_1$ and $X_2$ were correlated (Table \ref{table:1}b). For example, for $(\beta_1 ,\beta_2)=(1-1)$, $(\zeta_1, \zeta_2)=(1,-1)$, and $\rho=0.5$, the ratio of the variance of $(\hat \beta_1, \hat \beta_2)$ based on $500$ incident cases and controls alone was three times larger compared to the variance after adding $n_2=1000 $ prevalent cases (Figure 2b, and Supplementary Table S6). Additional results are given in Supplementary Tables S1-S5.

\begin{table}[ht!]
\centering
\begin{minipage}{150mm}
\caption{Estimation of the log odds ratios ($\beta_1, \beta_2$) and survival parameters ($\kappa_1, \kappa_2, \zeta_1, \zeta_2$) when the sample size of the prevalent cases varies: $n_2 = 0, 500, 1000$, with $n_0=n_1=500$. The true log odds ratios were $\beta = (0, 0)$ for (a) and $\beta = (1, -1)$ for (b). For both (a) and (b), data were generated with $\rho = 0.5$ and $\xi = 25$. Estimates (Est), empirical standard deviations (SD$_{\text{emp}}$) and standard deviation estimates based on the asymptotic covariance matrix (SD$_{\text{asy}}$) are based on 1000 replications of the simulation.}{%
	\begin{tabular}{lcccccccc}
		\toprule
		%\multirow{2}{*}{(a)} 
		(a) & $\alpha^*$ & $\nu^*$ & $\beta_1$ = 0 & $\beta_2$ = 0 & $\kappa_1$ = 1 & $\kappa_2$ = 1 & $\zeta_1$ = 1 & $\zeta_2$ = -1\\
		\cmidrule{2-9}
		& \multicolumn{8}{c}{$n_0=500, n_1=500, n_2 = 0$}\\
		Est & 0.000 &  & 0.001 & -0.002 &  &  &  &  \\ 
		SD$_{\text{asy}}$ & 0.004 &  & 0.073 & 0.074 &  &  &  &  \\ 
		SD$_{\text{emp}}$ & 0.003 &  & 0.073 & 0.073 &  &  &  &  \\ 
		\\
		& \multicolumn{8}{c}{$n_0 = 500, n_1$ = 500, $n_2$ = 500}\\%
		Est & 0.001 & -0.473 & 0.002 & 0.001 & 1.017 & 1.011 & 1.021 & -1.019 \\ 
		SD$_{\text{asy}}$ & 0.004 & 0.105 & 0.068 & 0.068 & 0.093 & 0.141 & 0.100 & 0.100 \\ 
		SD$_{\text{emp}}$ & 0.002 & 0.101 & 0.066 & 0.068 & 0.094 & 0.138 & 0.105 & 0.103 \\
		\\ 
		& \multicolumn{8}{c}{$n_0 = 500, n_1$ = 500, $n_2$ = 1000}\\%

		Est & 0.001 & 0.223 & 0.000 & -0.000 & 1.007 & 1.003 & 1.007 & -1.008 \\ 
		SD$_{\text{asy}}$ & 0.003 & 0.077 & 0.066 & 0.066 & 0.065 & 0.101 & 0.071 & 0.071 \\ 
		SD$_{\text{emp}}$ & 0.002 & 0.076 & 0.069 & 0.065 & 0.065 & 0.100 & 0.070 & 0.072 \\ 
		\\
		\hline
		\vspace{0.01in} \\
		(b) & $\alpha^*$ & $\nu^*$ & $\beta_1$ = 1 & $\beta_2$ = -1 & $\kappa_1$ = 1 & $\kappa_2$ = 1 & $\zeta_1$ = 1 & $\zeta_2$ = -1\\
		\cmidrule{2-9}
		& \multicolumn{8}{c}{$n_0=500, n_1=500, n_2 = 0$}\\
		Est & -0.504 &  & 1.006 & -1.010 &  &  &  &  \\ 
		SD$_{\text{asy}}$ & 0.051 &  & 0.089 & 0.089 &  &  &  &  \\ 
		SD$_{\text{emp}}$ & 0.049 &  & 0.089 & 0.087 &  &  &  &  \\ 
		\\
		& \multicolumn{8}{c}{$n_0 = 500, n_1$ = 500, $n_2$ = 500}\\%
		Est & -0.501 & -0.000 & 1.004 & -1.001 & 1.017 & 1.014 & 1.020 & -1.019 \\ 
		SD$_{\text{asy}}$ & 0.038 & 0.095 & 0.071 & 0.071 & 0.087 & 0.127 & 0.097 & 0.097 \\ 
		SD$_{\text{emp}}$ & 0.037 & 0.094 & 0.070 & 0.073 & 0.090 & 0.127 & 0.100 & 0.099 \\ 
		\\
		& \multicolumn{8}{c}{$n_0 = 500, n_1$ = 500, $n_2$ = 1000}\\%
		Est & -0.500 & 0.697 & 1.002 & -1.002 & 1.007 & 1.004 & 1.008 & -1.007 \\ 
		SD$_{\text{asy}}$ & 0.032 & 0.067 & 0.066 & 0.066 & 0.060 & 0.090 & 0.068 & 0.068 \\ 
		SD$_{\text{emp}}$ & 0.032 & 0.066 & 0.067 & 0.065 & 0.061 & 0.089 & 0.070 & 0.071 \\ 
		\bottomrule
\end{tabular}}
\end{minipage}
\label{table:1}
\end{table}

\begin{figure}
\centerline{\includegraphics[width=5in]{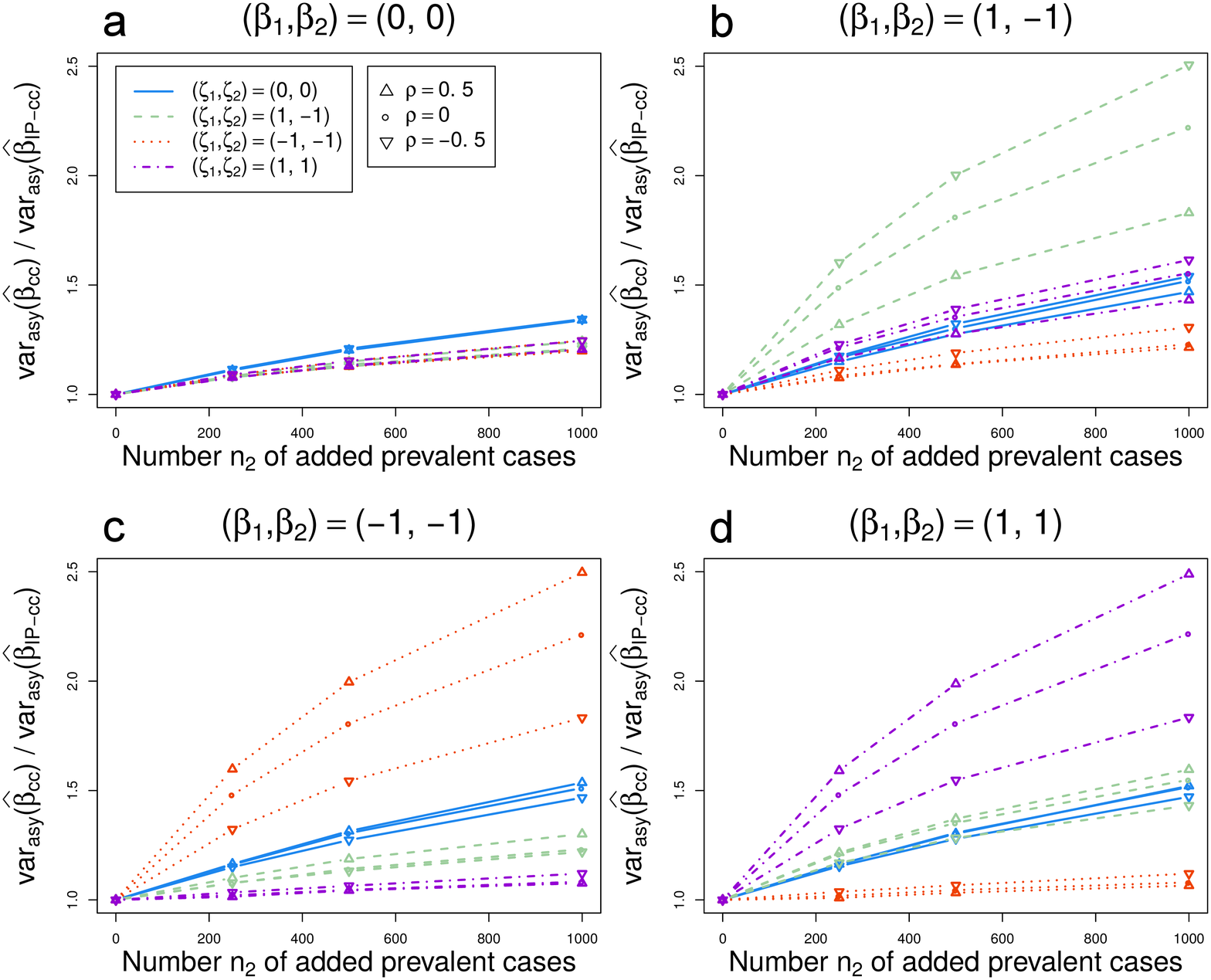}}
\caption{Efficiency of log odds ratio estimates when $n_2=250, 500 $ or $1000$  prevalent cases (shown on the x-axis) are  added to  $n_0 = 500$ controls and  $n_1 = 500$ incident cases (denoted by  $\hat{\beta}_{IP-cc}$), compared to
	those   estimated from controls and incident cases only (denoted by $\widehat{\beta}_{cc}$).  Asymptotic variances are used for both estimates.  The ratios of the variance estimates are based on 1000 replications of the simulation.}
\label{fig:sim2b}
\end{figure}

\subsection{Increasing the proportion of prevalent cases}
We next examined the efficiency of estimates $\hat \beta$ 
when the total number of cases was fixed at $n_1 + n_2=500$, but the proportion of prevalent cases increased, $n_2/(n_1 + n_2) = 0, 0.2, 0.5, 0.8$ or $1.0$. The number of controls was $n_0 = 500$.

Results in Figure 3 show that replacing incident with prevalent cases resulted in an appreciable loss of efficiency of estimates $\hat \beta$ in most settings. This was especially apparent when $\beta=0$, where all the ratios var($\hat \beta_{cc}$)/var($\hat \beta_{IP-cc}$) were below one. However, similar to the results in Section \ref{sec:adding}, when $\beta$ and $\zeta$ had the same sign and magnitude, there was a gain in efficiency of $\hat \beta$ when prevalent, instead of incident, cases were used. For example, for $\beta = (1, -1)$ and $\zeta=(1,-1)$ with $\rho=-0.5$, the SD$_{\text{emp}}$ for $\widehat{\beta}$'s decreased from 0.107 to 0.091 as the proportion of prevalent cases increased from 0 to 100\%, resulting in a 28\% efficiency gain, as measured by the ratio of the corresponding variances (Supplementary Table S9 and additional results in Supplementary Tables S7-S13).

\subsection{Efficiency of $\hat \beta$ for added prevalent versus incident cases}

When designing a study, an investigator may have the choice of including additional incident or additional prevalent cases, possibly associated with different costs. We thus further investigated the difference in efficiency of $\hat \beta$ when adding either incident or prevalent cases to a study comprised of a ``base sample'' of $500$ controls and $500$ incident cases. We first added from 20 to 1000 incident cases in increments of 20 to the base sample, and estimated var$_{\text{asy}}(\hat \beta_{cc})$. %, for $n_1=500$ to $1500$ incident cases. 
Then, for each value of var$_{\text{asy}}(\hat \beta_{cc})$, we found the number $n_2$ of prevalent cases that, if added to the $n_0=500$ controls and $n_1 = 500$ incident cases, resulted in $0.9985 \le \mbox{var}_{\text{asy}}(\hat \beta_{IP-cc}) / \mbox{var}_{\text{asy}}(\hat \beta_{cc}) \le 1.0015$. We also increased $n_2$ in increments of 20.

Figure \ref{fig:sim2c} shows the relationship between additional incident and additional prevalent cases for all scenarios considered in Section 4.2. For most settings, using prevalent cases led to less efficient estimates of $\beta$, indicated by lines above the 45$^\circ$ (gray dot-dashed) line, that corresponds to equal variance for the same number of added incident or prevalent cases. This loss of efficiency was particularly apparent when $\beta_1=\beta_2=0$, where even for $\zeta_1=\zeta_2=0$, approximately $n_2=300$ prevalent cases yielded the same variance of $\hat \beta$ as 200 additional incident cases. However, when $\beta$ and $\zeta$ had the same sign and magnitude, a prevalent case 
provided more information 
than an additional incident case, as indicated by the lines below the 45$^\circ$ line. For example, for $\beta=(-1,-1)$ and $(\zeta_1,\zeta_2)=(-1,-1)$, using $n_2=200$ prevalent cases resulted in the same variance of $\hat \beta$ as adding 400 incident cases to the base study sample.

\subsection{Robustness to mis-specification of the survival model} 

As our method requires specifying a parametric survival distribution $S$ to model the backward time,
we examined the robustness of the method to misspecification of $S$.

First, we studied the estimates of $\beta$ when the backwards time was generated using $S$ that had a proportional hazards form with a Weibull baseline and was a function of three covariates $X_1, X_2, X_3$, but we omitted $X_3$ in fitting the model. When $X_3$ was uncorrelated with $X_1$ and $X_2$, all parameter estimates were unbiased and there was virtually no loss of efficiency (data not shown). When $X_3=0.5 X_1+0.5 X_2 + \epsilon, \epsilon \sim N(0,0.25)$, there was appreciable bias in the parameter estimates of $S$, but no noticeable bias in $\hat \beta$. For example, when $(\beta_1, \beta_2)=(1,-1)$ and $n_2=1000$ prevalent cases were added to $500$ incident cases and $500$ controls, the log odds ratio estimates were $\hat \beta_1=1.003$ and $\hat \beta_2=-1.002$. However, $\hat \zeta_1 = 0.373$ and $\hat \zeta_2 = -1.117$ instead of $(\zeta_1,\zeta_2)=(1,-1)$ (Supplementary Table S14). There was no noticeable impact of the survival model misspecification on the efficiency of estimates of $\beta$. 

Assuming a proportional hazards model for $S$, we also assessed the robustness of our method to misspecification of the baseline hazard of $S$.
We simulated the backward time for prevalent cases with a Weibull baseline hazard with shape $\kappa_1 = 3$ and scale $\kappa_2= 25$ or a piecewise-constant baseline hazard with $\lambda_{01}$ = 0.025, $\lambda_{02}$ = 0.1, $\lambda_{03}$ = 0.25 with breakpoints at $\tau_1$ = 0, $\tau_2$ = 10, $\tau_3$ = 30, $\tau_4 > 45$, 
and fit the IP-case-control likelihood (\ref{loglik}) using either a Weibull or piecewise-constant baseline hazard.
We generated the data with $\beta = (0, 0)$ or $(1, -1)$ and $\zeta = (0, 0)$ or $(1, -1)$ with sample sizes $n_0=n_1=n_2=500$. 
Estimates of $\beta$ were virtually unbiased, despite  misspecification of the baseline hazard,
as were estimates for the $\zeta$
(Supplementary Table S15). For this simulation the data were generated prospectively, as described in Supplementary Section 4.4.1.

\begin{figure}
	% The arguments in the next line are {height}{optional width}{used only by OUP for typesetting}[filename, in directory art]
	\centerline{\includegraphics[width=5in]{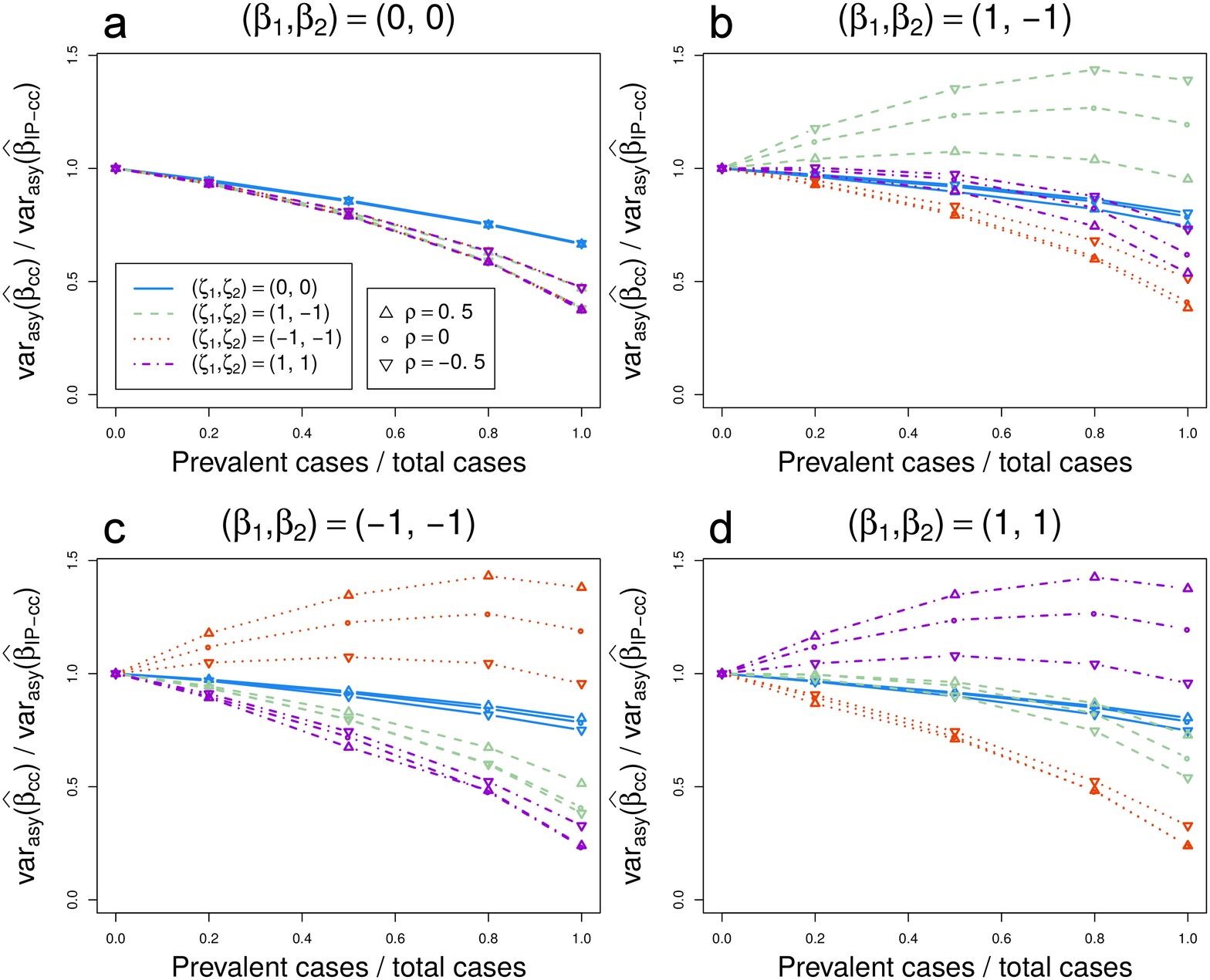}}
	% note that files may not be rotated
	\caption{Efficiency  of    $\hat{\beta}_{IP-cc}$  compared to an estimate based on controls and incident cases only ($\widehat{\beta}_{cc}$) as the proportion of prevalent cases out of the total number of  cases, $n_2/(n_1 + n_2)$ (on the x-axis) increases, for  fixed $n_1 + n_2= 500$, and $n_0=500$ controls.  Asymptotic variances are used for both estimates.  The ratios of the variance estimates are based on 1000 replications of the simulation.}
	\label{fig:sim2a}
\end{figure}

\begin{figure}
	% The arguments in the next line are {height}{optional width}{used only by OUP for typesetting}[filename, in directory art]
	\centerline{\includegraphics[width=5in]{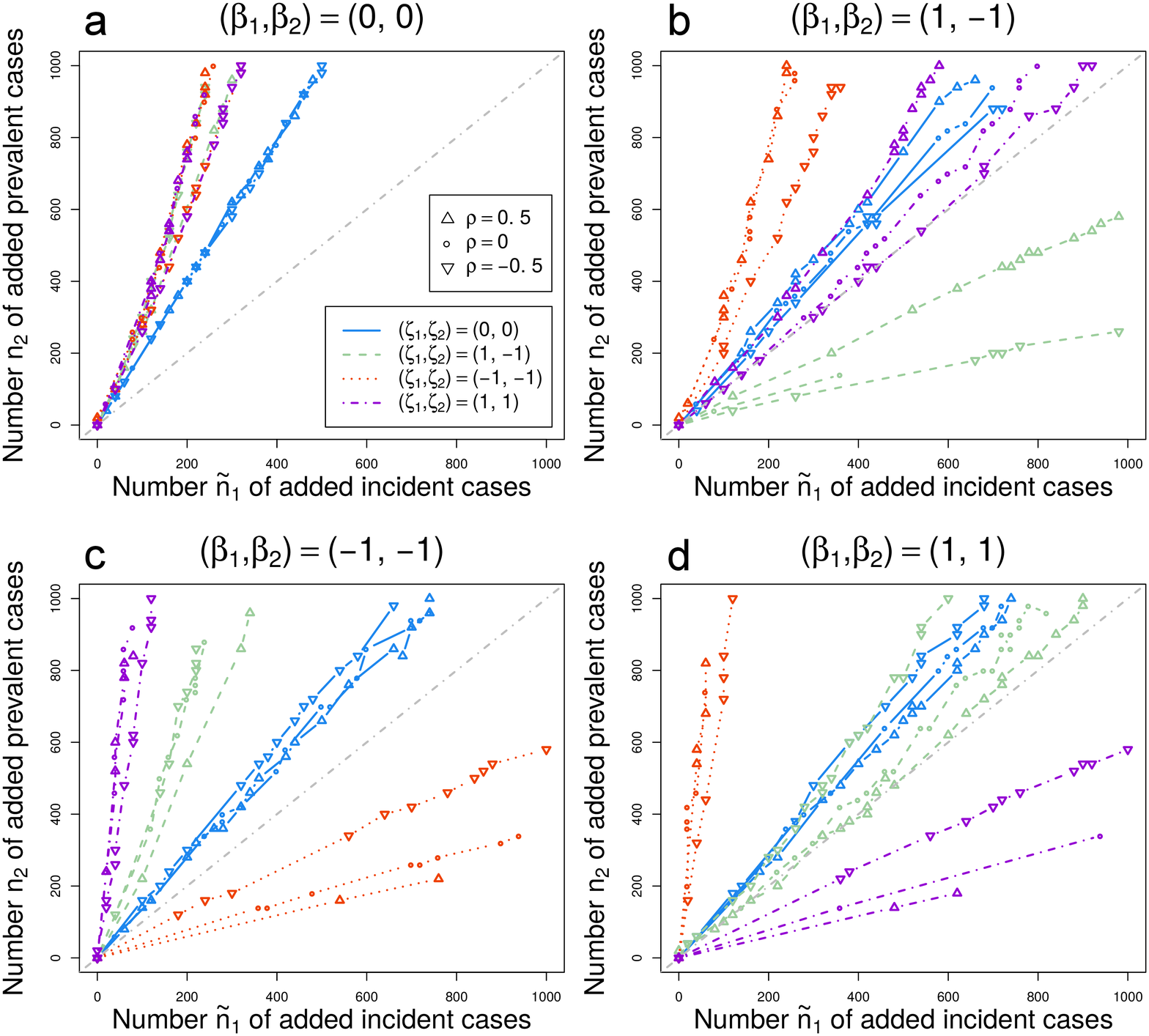}}
	% note that files may not be rotated
	\caption{Number $n_2$ of    prevalent cases on the  y-axis, that when added to     $n_0 = 500$ controls and  $n_1 = 500$ incident cases 
		yields the same efficiency of $\widehat{\beta}_{IP-cc}$ as 
		$\widehat{\beta}_{cc}$ when $\tilde n_1$ additional  incident cases (on the x-axis) are  added to  $n_0 = 500$ controls and  $n_1 = 500$ incident cases. Efficiency $=\mbox{var}_{\text{asy}}(\widehat{\beta}_{IP-cc})/\mbox{var}_{\text{asy}}(\widehat{\beta}_{cc})$. 
		%Variance estimates are based on sample sizes of $n_0 = 500$, $n_1 = 500$ plus an additional number of prevalent cases, var$_{\text{asy}}(\widehat{\beta}_{IP-cc})$, (x-axis) or incident cases, var$_{\text{asy}}(\widehat{\beta}_{cc})$, (y-axis).
		The variance estimates are based on 500 replications of the simulation.}
	\label{fig:sim2c}
\end{figure}

\section{Data example}

We now illustrate our model using data from the study that motivated this work,
a case-control study conducted within the United States Radiologic Technologists Study to assess associations of single nucleotide polymorphisms (SNPs) in candidate genes with risk of female breast cancer \citep{bhatti08a}.
This study used information from the first two surveys, conducted between 1984-1989, 1993-1998. Incident cases were women who answered both surveys and were diagnosed with a primary breast cancer between the two surveys.
Eligible controls were women who at the age of diagnosis of a case were were breast cancer free. Controls were  frequency matched to cases  by year of birth in five year strata.
 Prevalent cases were women who answered only one of the two surveys and who reported a prior breast cancer diagnosis. Their backward time was defined as the difference between the year of the survey and the year of their diagnosis. All breast cancer diagnoses were confirmed based on pathology or medical records.

The covariates used in our analysis were age at diagnosis for cases or age at selection for controls (in categories: $\le22, (22, 40], (45, 50], (50, 55], >55$); the year when the woman started working as a radiation technologist (1 if $\le$1955, 0 if $>1955$); smoking status (1 if former/current, 0 if never); history of breast cancer among first degree relatives (yes/no); BMI (kg/m$^2$) during their 20's (in categories: $\le20$, (20-25], $>25$); BMI during 20's among women diagnosed after 50 years of age (coded as BMI during 20's for women diagnosed at $\ge50$, and 0 otherwise, to capture the age dependent effect of BMI on breast cancer risk); history of heart disease (yes/no); alcohol consumption (1 if $\ge7$ drinks/week, 0 otherwise); and genotype for three SNPs: rs2981582 (1 if TC/TT, 0 if CC); rs889312 (1 if CA/CC, 0 if AA); and rs13281615 (1 if GG/GA, 0 if AA). We restricted our analysis to women with complete covariate information, and used data from 663 controls, 345 incident cases, and 213 prevalent cases.

The demographic characteristics differed between controls, incident and prevalent cases (Supplementary Table S16). The prevalent cases were older than incident cases and controls, more likely to have started work as a radiation technologist before 1955, more likely to be current smokers, and to have a first-degree relative with breast cancer. 

We compared log odds ratio estimates $\hat \beta$ from the following models: (A) \emph{Incident model}: standard logistic regression model fit to incident cases only and controls; (B) \emph{Na\"ive model}: standard logistic regression fit to controls and incident plus prevalent cases combined without accounting for survival bias in the prevalent cases; (C) \emph{IP-case-control}: IP-case-control likelihood (\ref{loglik}) fit to incident cases, controls, and prevalent cases accounting for survival bias.

The covariates in the logistic models were rs2981582, rs889312, rs13281615, age at diagnosis or selection, year first worked, family history, BMI in 20's, BMI in 20's (50+), and alcohol consumption. The survival sub-model in (C) was a Cox proportional hazards model with a Weibull baseline hazard, with the same covariates as in the logistic sub-model plus smoking status and history of heart disease. 
The support of the backward time was 0 to $\xi = 40$, where $\xi$ was chosen to be larger than the maximum backward time (35 years) among the prevalent cases. We computed jackknife standard errors (SEs) 
and also used them to compute 95\% confidence intervals (CIs), assuming normality of the log odds ratio (log(OR)) or log hazard ratio (log(HR)) estimates.

Results of the analysis are summarized in Table \ref{table:usrt-weib-40}. For model (A), the following covariates were significantly associated with breast cancer incidence (95\% CIs are given in parentheses):
SNP rs13281615, log(OR) = 0.40 (0.12, 0.69), year first worked, log(OR) = -0.84 (-1.23, -0.45), family history of breast cancer, log(OR) = 0.54 (0.25, 0.83), and BMI in ones 20's, log(OR) = -0.34 (-0.60, -0.08). For model (B), the significant covariates were: SNP rs13281615, log(OR) = 0.34 (0.10, 0.58), family history of breast cancer, log(OR) = 0.57 (0.32, 0.83), and BMI in ones 20's, log(OR) = -0.37 (-0.59, -0.14). Nearly all estimates from model B were attenuated compared to those from model A. 

For model (C), the following covariates were associated with breast cancer incidence:  SNP rs13281615, log(OR) = 0.32 (0.05, 0.58), year first worked, log(OR) = -0.34 (-0.65, -0.03), family history of breast cancer, log(OR) = 0.53 (0.27, 0.79), and BMI in ones 20's, log(OR) = -0.34 (-0.57, -0.11). The log(OR) estimates in the IP-case-control model for rs981782, rs889312 and BMI in 20's were close to those estimated from the incident model, with smaller standard errors. The log(OR) estimates for age at diagnosis and year first worked were somewhat lower than the estimates of model (A) (Table \ref{table:usrt-weib-40}). However, those two variables were the ones significantly associated with the backward time, with log(HR) = 0.48 (0.28, 0.68) for age at diagnosis, log(HR) = -1.48 (-2.03, -0.93) for year first worked. Not surprisingly, the baseline hazard increased with increasing backward time.

Based on the likelihood ratio test, using an asymptotic $\chi^2_6$ cutoff value, the IP-case control model with the three SNPs in the logistic and the survival models fit the data significantly better than a model without the SNPs (p = 0.033).

\begin{table}
\caption{Estimated log odds ratios (ORs) and log hazard ratios (HRs) and jackknife standard errors (SEs) for the association of rs2981582, rs889312 and rs13281615 adjusted for other potential risk factors from three models: \emph{Incident model}: incident cases and controls, estimates from a logistic model; \emph{Na\"ive model}: incident and prevalent cases combined and controls, estimates from a logistic model; \emph{IP-case-control model}: incident cases, prevalent cases and controls, estimates accounting for survival bias based on the likelihood equation (\ref{fullL}).}{%
\centering
	\begin{tabular}{lccc}
		\\
		\toprule
		& Incident (A) & Na\"ive (B) & IP-case-control (C) \\
		& $n_I/n_C = 345/663$ & $n_{I+P}/n_C = 558/663$ & $n_I/n_P/n_C = 345/213/663$\\
		& log(OR) (SE) & log(OR) (SE) & log(OR) (SE) \\
		\cmidrule{2-4}
		% from 170915-analysis-case-org-1yr-3snps-indiv-snps-jack
		\\
		%FGFR2 rs2981582 & 0.170 (0.144) & 0.122 (0.123) & 0.137 (0.134) \\ 
		%MAP3K1 rs889312 & 0.228 (0.137) & 0.188 (0.118) & 0.237 (0.127) \\ 
		%POU5F1P1 rs13281615 & 0.404 (0.147) & 0.341 (0.124) & 0.317 (0.134) \\ 
		rs2981582 & 0.170 (0.144) & 0.122 (0.123) & 0.137 (0.134) \\ 
		rs889312 & 0.228 (0.137) & 0.188 (0.118) & 0.237 (0.127) \\ 
		rs13281615 & 0.404 (0.147) & 0.341 (0.124) & 0.317 (0.134) \\ 
		Age at diagnosis/selection & 0.133 (0.073) & -0.056 (0.060) & 0.059 (0.063) \\ 
		Year first worked & -0.838 (0.198) & 0.022 (0.151) & -0.341 (0.159) \\ 
		Family history & 0.542 (0.148) & 0.574 (0.128) & 0.527 (0.133) \\ 
		BMI in 20s & -0.341 (0.132) & -0.366 (0.113) & -0.342 (0.118) \\ 
		BMI in 20s (50+) & 0.226 (0.213) & 0.221 (0.184) & 0.162 (0.199) \\ 
		7+ alcoholic drinks/week & 0.131 (0.203) & 0.109 (0.173) & -0.004 (0.196) \\ 
		\\
		& & & log(HR) (SE) \\
		\\
		%\cmidrule{4}
		%FGFR2 rs2981582 & & & 0.054 (0.234) \\ 
		%MAP3K1 rs889312 & & & 0.210 (0.194) \\ 
		%POU5F1P1 rs13281615 & & & -0.122 (0.223) \\ 
		rs2981582 & & & 0.054 (0.234) \\ 
		rs889312 & & & 0.210 (0.194) \\ 
		rs13281615 & & & -0.122 (0.223) \\ 
		Age at diagnosis/selection & & & 0.483 (0.103) \\ 
		Year first worked & & & -1.482 (0.280) \\ 
		Ever smoker & & & -0.136 (0.165) \\ 
		Family history & & & -0.188 (0.153) \\ 
		BMI in 20s & & & 0.094 (0.158) \\ 
		BMI in 20s (50+) & & & -0.240 (0.314) \\ 
		History of heart disease & & & 0.021 (0.406) \\ 
		7+ alcoholic drinks/week & & & -0.417 (0.350) \\
		\\
		$\kappa_1$, $\kappa_2^{\; b}$, Est (SE)& \multicolumn{3}{r}{ 1.581 (0.317), 11.147 (2.134)} \\
		\bottomrule
\end{tabular}}
\label{table:usrt-weib-40}
%\begin{tabnote}
$^a$ rs2981582: 1 if TC/TT, 0 if CC; rs889312 SNP: 1 if CA/CC, 0 if AA; rs13281615 SNP: 1 if GA/GG, 0 if AA; age at diagnosis/selection: coded with a trend based on categories $\le22, (22, 40], (45, 50], (50, 55], >55$; year first worked: 1 if $\le$1955, 0 otherwise; BMI (kg/m$^2$) in 20s: coded with a trend based on categories $\le 20, (20, 25], >25$; BMI in 20s (50+): coded as BMI in 20s among subjects with age at diagnosis of $\ge50$ years, 0 otherwise; ever smoker: 1 if current or former smoker, 0 otherwise; 7+ alcoholic drinks/week: 1 if $\ge7$ drinks per week, 0 otherwise.\\
$^b$ $\kappa_1$ and $\kappa_2$ are Weibull baseline hazard shape and scale parameters, respectively.
%\end{tabnote}
%\end{minipage}
\end{table}

\section{Discussion}

The distribution of exposures among prevalent cases, individuals who have a prior disease diagnosis and are alive at the time of sampling for a case-control study, differs from that among incident cases when the exposures are associated with survival after disease onset.
Thus  na\"ively combining  prevalent cases with incident cases in the analysis of case-control data without accounting for their survival bias leads to  biased estimates of log odds ratios for association \citep{begg87a}. 

In this paper we propose a semi-parametric model to incorporate data on covariates and the observed backward time from prevalent cases, 
to obtain unbiased estimates of exposure-disease association. 
We propose a three-group exponential tilting, or density ratio, model to accommodate   two case groups and one control group,  tht we  assume is  an appropriate comparison group for the  incident cases. 
We  provide a semi-parametric method for estimation based on empirical likelihood \citep{qin94a, qin98a}. 
%To ensure unbiased sampling of the control group,  incident-density sampling could be  used to obtain incident cases and controls (i.e. a nested case-control design within a cohort)

Many authors dealt with the issue of length-bias when estimating survival parameters based on a prevalent cohort \citep[e.g.][]{cook11a, huang12a, zhu17a}. However, very few publications use prevalent cases when samples are ascertained cross-sectionally. 
Without using any information on follow-up, \citet{chan13a} estimated the impact of a covariate on the survival distribution in a log-linear model by showing
that 
the covariate sampling distribution of prevalent cases compared to incident cases could be expressed using an exponential tilting model. 
To our knowledge only 
\citet{begg87a} addressed adjusting for survival bias when comparing prevalent cases to controls to estimate incidence odds ratios, again, not using any follow-up information. They modeled the backward time distribution based on an accelerated failure time model for survival and estimated the parameters using quasi-likelihood techniques. 
Incidence log odds ratio parameters were then estimated by subtracting a bias term from the log odds ratio estimates obtained from a standard logistic model fit to controls and prevalent cases. 

In contrast to the approach by \citet{begg87a}, we propose a semi-parametric likelihood that yields root $N$ consistent and fully efficient estimates of the incident log odds parameters. 
We show that the corresponding likelihood ratio statistic has a 
standard asymptotic chi-square distribution, which makes the test easy to use and therefore  relevant for practical applications. 
Based on simulations,  the efficiency gains or losses when prevalent cases are added to, or used instead of, incident cases depend on the ratio of the incident to prevalent cases,  and the correlation structure among the covariates in the incidence and survival sub-models.
%and the parametric model for the backward time. 
Surprisingly, in some settings, prevalent cases were more informative than incident cases, which warrants further investigation in future work. 

A limitation of our approach is that the model for the backward time is fully parametric. However, based on simulations, the estimates of the log odds ratios were not affected by reasonable misspecification of the model for the backward time. 
Our method is thus very appealing in settings where there is little concern about recall bias for the main exposure and the number of available incident cases is limited.

\section*{Supplementary material}
\label{SM}
Web Appendices referenced in Sections 3, 4.2, 4.3, 4.5, 5, and Appendix 2, include proofs of equations (17) and (20), %(B1) and (B4), 
calculation of $\Sigma$ in equation (17), and additional simulation results including misspecified models, are available with this paper at the Biometrics website on Wiley Online Library.

\section*{Acknowledgement}
The authors thank Michele Doody at the National Cancer Institute for providing the data, Jerry Reid at the American Registry of Radiologic Technologists, Diane Kampa and Allison Iwan at the University of Minnesota, Jeremy Miller and Laura Bowen at Information Management Services, and the radiologic technologists who participated. This work utilized the computational resources of the NIH HPC Biowulf cluster (http://hpc.nih.gov).

\bibliographystyle{biom}
\bibliography{references-prev-cc}

\appendix
\section{Appendix 1}

\subsubsection*{Derivation of the profile log-likelihood (\ref{loglik})$\;\;$} 
Let $w_1(x) = \exp(\alpha^*+x \beta)$ and $w_2(x) = \exp\{\nu^*+x \beta+\log \mu(x,\gamma)\}$ where $\alpha^*$ and $\nu^*$ are defined as in equations (\ref{eq:tilt1}) and (\ref{eq:f2x}), respectively. We then rewrite the likelihood in (\ref{fullL}) as 
{\footnotesize \begin{eqnarray*}
		\mathcal{L}
		&=& \prod_{i=1}^N f_0(x_i) \prod_{i=n_0+1}^{n_0+n_1}w_1(x_i) 
		\prod_{i=n_0+n_1+1}^{N} \left\{ w_2(x_i) \frac{S(a_i\mathbin{|}x_i, \gamma)}{\mu(x_i, \gamma)}\right\}.
\end{eqnarray*}} 
Following \cite{qin98a}, we estimate $p_i=f_0(x_i) = P(X=x_i)$, $i=1,\ldots,N$, empirically under the following constraints: \mbox{(1) $\sum_{i=1}^Np_i = 1, p_i \geq 0$}, (2) $\sum_{i=1}^N p_i w_1(x_i)$, %\exp(\alpha^*+x_i\beta) = 1$, 
(3) $\sum_{i=1}^Np_i w_2(x_i)$, %\exp\{\nu^*+x_i\beta+\log\mu(x_i, \gamma)\} = 1$, 
by accommodating them in the log-likelihood $\log(\mathcal{L})$ using Lagrange multipliers, $\lambda_i$, $i=0, 1, 2$. The $p_i$'s and $\lambda_i$'s are explicitly computed by maximizing the constrained log-likelihood: %All parameters are then estimated by maximizing 
\begin{eqnarray*}
	\ell_c &=&\sum_{i=1}^N\log p_i + \sum_{i=n_0+1}^{n_0+n_1}\log w_1(x_i)+\sum_{i=n_0+n_1+1}^{N} \log w_2(x_i)\left\{\frac{S(a_i\mathbin{|}x_i, \gamma)}{\mu(x_i, \gamma)}\right\}\\ &\;& + \, \lambda_0 (1-\sum_{i=1}^N p_i) + N \lambda_1 \left\{1- \sum_{i=1}^{N}p_iw_1(x_i)\right\} + N \lambda_2 \left\{1- \sum_{i=1}^{N}p_i w_2(x_i)\right\}.
\end{eqnarray*}

Taking derivatives with respect to the $p_i$'s we explicitly compute $\lambda_0$ and $p_i$. 
\begin{equation}
	\label{eq:partial-lc}
	\frac{\partial \ell_c}{\partial p_i} = \frac{1}{p_i}-\lambda_0-N \lambda_1 w_1(x_i)-N \lambda_2 w_2(x_i) = 0 
\end{equation}
and
{\small \begin{eqnarray}
		\sum_{i=1}^N p_i\frac{\partial \ell_c}{\partial p_i} &=& \sum_{i=1}^Np_i\frac{1}{p_i}-\sum_{i=1}^Np_i\lambda_0-\sum_{i=1}^Np_iN \lambda_1 w_1(x_i) \nonumber-\sum_{i=1}^N p_i N \lambda_2 w_2(x_i) \nonumber\\
		&=& N-\lambda_0-N\lambda_1-N\lambda_2 = 0 %\nonumber \\
		\Rightarrow \lambda_0 = N(1-\lambda_1-\lambda_2). \label{lambda0}
\end{eqnarray}}
Plugging (\ref{lambda0}) into equation (\ref{eq:partial-lc}) yields
\begin{equation}
	\label{pi}
	p_i = \frac{1}{N \left[1 + \lambda_1\{w_1(x_i)-1\} + \lambda_2\{w_2(x_i)-1\}\right]} \\
\end{equation}
and the profile log-likelihood for the remaining parameters $ \ell_p(\lambda_1,\lambda_2,\alpha^*,\nu^*,\beta,\gamma)$, is
\begin{eqnarray}
	\label{eq:ellp}
	\ell_p(\lambda_1,\lambda_2,\alpha^*,\nu^*,\beta,\gamma)
	&=& -\sum_{i=1}^N\log\left[1+\lambda_1\{w_1(x_i)-1\}+\lambda_2\{w_2(x_i)-1\}\right]\\
	&\;& +\,\sum_{i=n_0+1}^{n_0+n_1}\log w_1(x_i)+\sum_{i=n_0+n_1+1}^{N} \log w_2(x_i)\left\{\frac{S(a_i\mathbin{|}x_i, \gamma)}{\mu(x_i, \gamma)}\right\} + c_1\nonumber
\end{eqnarray}
where $c_1 = -N\log(N)$.
Differentiation of $ \ell_p(\lambda_1,\lambda_2,\alpha^*,\nu^*,\beta,\gamma)$ %with respect to the remaining parsmeters 
yields 
%Lagrange multipliers $\lambda_1$ and $\lambda_2$ and $\alpha^*,\nu^*, \beta$ and $\gamma$ yields 
\begin{equation}
	\label{dlpdl1}
	\frac{\partial \ell_p}{\partial \lambda_k} = \sum_{i=1}^N\frac{w_k(x_i)-1}{1+\lambda_1\{w_1(x_i)-1\}+\lambda_2\{w_2(x_i)-1\}}=0, k=1,2.
\end{equation}
%\begin{equation}
%\label{dlpdl2}
%\frac{\partial \ell_p}{\partial \lambda_2} =\sum_{i=1}^N\frac{w_2(x_i)-1}{1+\lambda_1\{w_1(x_i)-1\}+\lambda_2\{w_2(x_i)-1\}}=0.
%\end{equation}
and 
%{\footnotesize 
\begin{eqnarray}
	\label{dldpalpha}
	\frac{\partial \ell_p}{\partial \alpha^*}&=&
	-\sum_{i=1}^N\frac{\lambda_1 w_1(x_i)}{1+\lambda_1\{w_1(x_i)-1\}+\lambda_2\{w_2(x_i)-1\}} + n_1=0 \\
	\label{dldpnu}
	\frac{\partial \ell_p}{\partial \nu^*}&=&
	-\sum_{i=1}^N\frac{\lambda_2w_2(x_i)}{1+\lambda_1\{w_1(x_i)-1\}+\lambda_2\{w_2(x_i)-1\}} + n_2=0 \\
	\frac{\partial \ell_p}{\partial \beta}&=&
	-\sum_{i=1}^N\frac{\lambda_1x_iw_1(x_i)+\lambda_2x_iw_2(x_i)}{1+\lambda_1\{w_1(x_i)-1\}+\lambda_2\{w_2(x_i)-1\}} + \sum_{i=n_0+1}^{n_0+n_1}x_i+\sum_{i=n_0+n_1+1}^{N}x_i=0 \nonumber \\
	\frac{\partial \ell_p}{\partial \gamma} &=&
	-\sum_{i=1}^N\frac{\lambda_2w_1(x_i)\frac{\partial}{\partial\gamma}\mu(x_i, \gamma)}{1+\lambda_1\{w_1(x_i)-1\}+\lambda_2\{w_2(x_i)-1\}} + \sum_{i=n_0+n_1+1}^{N}\frac{\partial}{\partial\gamma}\log S(a_i\mathbin{|}x_i, \gamma) =0. \nonumber
\end{eqnarray} %}

Next, we solve for $\lambda_1$ and $\lambda_2$. From equation (\ref{pi}) and the constraint that $\sum_{i=1}^N p_i = 1$ it follows that 
{\small \begin{eqnarray}
		\label{sums-to-n}
		\sum_{i=1}^N\frac{1}{1 + \lambda_1 \{w_1(x_i)-1\} + \lambda_2\{w_2(x_i)-1\}} = N 
\end{eqnarray}}
From equations (\ref{dlpdl1}) and (\ref{sums-to-n}) we have
{\small \begin{eqnarray}
		\label{sum1}
		\sum_{i=1}^N\frac{w_k(x_i)}{1+\lambda_1\{w_1(x_i)-1\}+\lambda_2\{w_2(x_i)-1\}} = N, \quad k=1,2.
\end{eqnarray}}
Then, using (\ref{dldpalpha}) and (\ref{sum1})
\[\frac{\partial \ell_p}{\partial \alpha^*}= 0 \Rightarrow \sum_{i=1}^N\frac{\lambda_1w_1(x_i)}{1+\lambda_1\{w_1(x_i)-1\}+\lambda_2\{w_2(x_i)-1\}} = n_1 \Rightarrow \widehat{\lambda}_1 = \frac{n_1}{N}\]
and similarly, from (\ref{dldpnu}) and (\ref{sum1}) for $\nu^*$ we get that $\widehat{\lambda}_2 = n_2/ N$. 
Plugging the estimates $\widehat{\lambda}_1$ and $\widehat{\lambda}_2$ into equation (\ref{eq:ellp}) yields
the profile likelihood in (\ref{loglik}).

\section{Appendix 2}

\subsubsection*{Proof of Theorem 1$\;\;$}

Using arguments similar to those in the proofs of Lemma 1 and Theorem 1, \citet{qin94a}, %Qin:1994
we have $\hat \theta -\theta = O_p(N^{-1/2})$.
Our proof begins by studying the behavior of $\ell_p(\theta)$ for $\theta = \theta_0+O_p(N^{-1/2})$, and 
we use the following Lemma.

\begin{lemma}
	\label{foundation}
	Assume that $\theta^{\T}=(\theta_{1}^{\T}, \theta_{2}^{\T}) $
	where $\theta_1$ and $\theta_2$ are $r$- and $s$-dimensional vectors, respectively.
	Let $\theta_0^{\T}=(\theta_{10}^{\T}, \theta_{20}^{\T})$ be its true value,
	and $\gamma =(\gamma_{1}^{\T}, \gamma_{2}^{\T})^{\T}
	= N^{1/2}(\theta-\theta_0)$ where $N$ is the sample size.
	Suppose for $\theta = \theta_0+O_p(N^{-1/2})$,
	it holds that
	\bas
	H(\theta) = C_N + a_N^{\T} \gamma - \frac{1}{2} \gamma^{\T}A\gamma + \varepsilon_N(\theta)
	\eas
	where $a_N=O_p(1)$, $A$ is a positive definite matrix, $C_N$ does not depend on $\theta$,
	and $\varepsilon_N(\theta)= o_p(1)$ for any fixed $\theta$.
	According to $\theta=(\theta_{1}^{\T}, \theta_{2}^{\T})^{\T}$,
	we partition $A$ into
	\[
	A = \left(
	\begin{array}{cc}
	A_{11}&A_{12}\\
	A_{21}&A_{22}\\
	\end{array}
	\right),
	\]
	and partition $a_N^{\T}$ into $(a_{N1}^{\T}, a_{N2}^{\T})$.
	If as $N\rightarrow \infty$, $a_N\convergeto N(0, A)$ in distribution, then
	\bit
	\item[(a)]
	the maximizer $\hat\theta$
	of $H(\theta)$ satisfies
	$ N^{1/2}(\hat \theta-\theta_0) = A^{-1}a_N + o_p(1) \convergeto N(0, A^{-1})$ in distribution,
	
	\item[(b)]
	\(
	2\{ \max_{\theta} H(\theta) - H(\theta_0) \} = a_N^{\T} A^{-1} a_N + o_p(1) \convergeto \chi_{r+s}^2
	\) in distribution, and
	\item[(c)]
	\(
	2\{ \max_{\theta} H(\theta)- \max_{\theta_2} H(\theta_{10}, \theta_2)\}
	= a_N^{\T} A^{-1} a_N - a_{N2}^{\T} A_{22}^{-1} a_{N2}+o_p(1)
	\convergeto \chi_{r}^2
	\) in distribution.
	\eit
\end{lemma}
\noindent 
Statement (a) in the above Lemma can be proven  by direct application of results from the {\it Basic Corollary} in \citet{Hjort:1993},
and statements (b) and (c) follow from statement (a).

%WAS \subsubsection{\blue Proof of Theorem 1 Result (1)}
\subsubsection*{Proof of Theorem 1 Result (1)$\;\;$}
Let $\xi=N^{1/2}(\theta-\theta_0)$ and
$\nabla_{\phi}$ denote the differentiation operator with respect to a generic parameter $\phi$.
After verifying that $\e\left\{ N^{-1 } \nabla_{\theta\theta^\T} \ell(\theta_0) \right\} = V \equiv (V_{ij})_{1\leq i, j\leq 4}$ (Supplementary Materials Section 3.2), %\ref{cal.sigma}).
by the second-order Taylor expansion, for $
u \equiv (u_1, u_2, u_3, u_4)^\T = N^{-1/2} \nabla_{\theta} \ell (\theta_0)
$, % we have
\bas
\ell(\theta)
&=&
C + u^{\T} \xi + \frac{1}{2} \xi^{\T} V \xi + o_p(1),
\eas
where $C$ is a constant not depending on $\theta$. 
As the maximum likelihood estimate $\hat \theta$ %MLE % $\hat \theta$ of $\theta$,
solves $\nabla_{\theta} \ell(\theta) = 0$, 
\bas
\hat \xi \equiv N^{1/2}(\hat \theta - \theta_0) = -V^{-1} u+o_p(1).
\eas
Note that each component of $u$ is a linear combination of sums of independent
and identically distributed random variables, therefore
by central limit theorem, $u$ has a limiting normal distribution.
In Section 1 of Supplementary Materials %\ref{proof-u}, 
we prove
\ba
\label{asy-u}
u \convergeto N(0, \Sigma)\;\; \mbox{ in distribution.}
\ea
Therefore $\hat \xi \convergeto N(0, \Omega)$ in distribution with
$\Omega = V^{-1} \Sigma V^{-1}$. This proves result (1) of Theorem 1.

\subsubsection*{Proof of Theorem 1 Results (2) and (3)$\;\;$}
For ease of exposition we partition $u = (u_a^\T, u_b^\T)^\T$ with
%\bas
$u_a=(u_1, u_2)^\T, u_b=(u_3, u_4)^\T
$ %\eas
and similarly partition $\xi$ and $V$ as
\bas
V=
\left(
\begin{array}{cc}
	V_{aa} & V_{ab} \\
	V_{ba} & V_{bb}
\end{array}
\right)
\eas
where
\bas
V_{aa}=
\left(
\begin{array}{cc}
	V_{11} & V_{12} \\
	V_{21} & V_{22}
\end{array}
\right),\quad
V_{ab}=V_{ba}^\T=
\left(
\begin{array}{cc}
	V_{13} & V_{14} \\
	V_{23} & V_{24}
\end{array}
\right),\quad
V_{bb}=
\left(
\begin{array}{cc}
	V_{33} & V_{34} \\
	V_{43} & V_{44}
\end{array}
\right)
\eas
Then the full likelihood function
\ba
\ell(\theta)
&=&
C + u^{\T} \xi + \frac{1}{2} \xi^{\T} V \xi + o(1) \nonumber \\
&=&
C + u_a^{\T} \xi_a + u_b^{\T} \xi_b + \frac{1}{2} \xi_a^{\T} V_{aa} \xi_a
+ \xi_a^{\T} V_{ab} \xi_b
+ \frac{1}{2} \xi_b^{\T} V_{bb} \xi_b + o_p(1).
\label{like}
\ea
To approximate the profile likelihood of $(\beta, \gamma)$,
we set $\nabla_{\xi_a}\ell(\theta) = 0$ and get
\ba
\label{app.xia}
\xi_a
= - V_{aa}^{-1}(u_a+V_{ab} \xi_b) + o_p(1).
\ea

Profiling out $(\alpha, \nu)$ from $\ell(\theta)$
or putting \eqref{app.xia} into \eqref{like} gives
\bas
\ell_p(\beta, \gamma)
&=&
C -\frac{1}{2}(u_a+V_{ab} \xi_b)^\T V_{aa}^{-1}(u_a+V_{ab} \xi_b) + u_b^{\T} \xi_b
+ \frac{1}{2} \xi_b^{\T} V_{bb} \xi_b + o_p(1) \\
&=&
C_1+ \frac{1}{2}\xi_b^\T(V_{bb} - V_{ba} V_{aa}^{-1} V_{ab}) \xi_b +
\xi_b^\T (u_b - V_{ba} V_{aa}^{-1}u_a ) + o_p(1)
\eas
where $C_1$ is another constant independent of $(\beta, \gamma)$.
This further implies
\begin{multline}
	2\{ \sup_{\beta, \gamma}\ell_p(\beta, \gamma) - \ell_p(\beta_0, \gamma_0) \} 
	=
	2\{ \sup_{\theta}\ell (\theta) - \sup_{\alpha, \nu} \ell (\alpha, \nu, \beta_0, \gamma_0) \} \\
	=
	- (u_b - V_{ba} V_{aa}^{-1}u_a )^{\T} (V_{bb} - V_{ba} V_{aa}^{-1} V_{ab})^{-1}(u_b - V_{ba} V_{aa}^{-1}u_a ) +o_p(1). \nonumber
\end{multline}

In section 2 of the Supplement we prove %\ref{proof-var}, we prove
\ba
\label{key.equality}
\var( u_b - V_{ba} V_{aa}^{-1}u_a )
= -(V_{bb} - V_{ba} V_{aa}^{-1} V_{ab}).
\ea
Then by Lemma 1, the likelihood ratio
$
2\{ \sup_{\beta, \gamma}\ell_p(\beta, \gamma) - \ell_p(\beta_0, \gamma_0) \}
$
converges in distribution to $\chi_p^2$, where $p$ is the dimension of $(\beta, \gamma)$.
Lemma 1 also indicates that the likelihood ratio test for any subvector of $(\beta, \gamma)$ 
has still a limiting central $\chi^2$ distribution. This proves results (2) and (3) of Theorem 1.

\label{lastpage}

\end{document}